\documentclass[11pt]{article}

\usepackage[final]{acl}

\usepackage{times}
\usepackage{latexsym}
\usepackage[T1]{fontenc}
\usepackage[utf8]{inputenc}
\usepackage{microtype}
\usepackage{inconsolata}
\usepackage{graphicx}
\usepackage{booktabs} 
\usepackage{amsmath}
\usepackage{xcolor}
\usepackage{booktabs}
\usepackage{multirow}
\usepackage{siunitx}
\usepackage[table]{xcolor}
\usepackage{algorithm}
\usepackage{algpseudocode}
\usepackage{amsmath}

\definecolor{sectiongray}{gray}{0.93}

\usepackage[most]{tcolorbox}
\usepackage{listings}
\usepackage{placeins}

\definecolor{promptblue}{RGB}{62, 94, 112}
\definecolor{promptbg}{RGB}{247, 250, 251}
\definecolor{prompttitle}{RGB}{62, 94, 112}
\definecolor{promptframe}{RGB}{62, 94, 112}
\definecolor{promptbg}{RGB}{247, 250, 251}

\newtcblisting{promptbox}[1]{
  enhanced,
  breakable,
  colback=promptbg,
  colframe=promptframe,
  coltitle=white,
  colbacktitle=prompttitle,
  title={#1},
  fonttitle=\bfseries,
  boxrule=0.7pt,
  arc=1.5mm,
  left=2mm,
  right=2mm,
  top=1mm,
  bottom=1mm,
  listing only,
  listing options={
    basicstyle=\ttfamily\scriptsize,
    breaklines=true,
    breakatwhitespace=false,
    columns=fullflexible,
    keepspaces=true,
    showstringspaces=false,
    upquote=true,
    tabsize=2
  }
}

\title{Rethinking Agentic RAG: Toward LLM-Driven \\ Logical Retrieval Beyond Embeddings}

\author{
Yuqi Zeng$^{1}$,
Qixiang Deng$^{1}$,
Yulei Wan$^{2}$,
Ruiquan Jiang$^{2}$,
Xiaoqing Zheng$^{1}$\thanks{Corresponding author.},
Xuanjing Huang$^{1}$ \\
$^{1}$School of Computer Science, Fudan University \\
$^{2}$Shanghai Kingstar Software Technology Co., Ltd. \\
\texttt{23307130067@m.fudan.edu.cn, zhengxq@fudan.edu.cn}
}

\begin{document}
\maketitle

\begin{abstract}

Recent advances in RAG have shifted toward an agentic paradigm, where LLMs interact with retrieval systems over multiple turns and iteratively refine queries based on intermediate results. 
At the same time, LLMs have demonstrated a strong ability to construct structured queries that precisely express their information needs. 
However, contemporary RAG systems remain heavily focused on engineering complex retrieval backends, including dense, hybrid, and graph-based retrieval architectures.
In this study, we argue that agentic RAG should delegate greater control to the LLM to steer the retrieval process, while relying on a lightweight retrieval interface that provides fine-grained control and faithfully executes the LLM's structured intent.
Guided by this principle, we propose an agentic RAG framework that enables LLMs to formulate retrieval intents using logical expressions while simplifying the retrieval backend to an inverted-index-based system. 
Extensive experiments show that our framework matches a strong agentic hybrid baseline, while substantially reducing construction and serving cost.
Moreover, we show that anchoring the retrieval process in logical queries substantially reduces hallucinations in generated responses.
\end{abstract}

\section{Introduction}

Retrieval-augmented generation (RAG) has become a fundamental framework for grounding the outputs of large language models (LLMs) in external knowledge \citep{NEURIPS2020_6b493230,izacard-grave-2021-leveraging}. 
More recently, advancements in RAG have shifted toward an agentic paradigm, wherein LLMs interact with retrieval backends over multiple turns~\citep{trivedi2023interleaving,ICLR2024_25f7be96,li2025search}.
This allows the model to progressively refine its queries to fulfill complex information needs until the required information is successfully retrieved or the model determines that the requested knowledge is unavailable.
Consequently, the effectiveness of modern RAG systems no longer depends solely on the quality of a single query and its retrieved results, but on the quality of multi-turn interactions between LLMs and retrieval backends. 
In particular, successful agentic RAG requires LLMs to accurately express retrieval intents, retrieval backends to faithfully execute those intents, and LLMs to adaptively refine subsequent queries based on intermediate results.

Retrieval backends are commonly constructed using either graph-based or embedding-based approaches.
Graph-based backends are typically motivated by the observation that complex queries often require decomposition into several successive sub-queries, particularly when the target information is scattered across disparate sources~\citep{edge2024local,gutierrez2024hipporag,zhu2025knowledge}. 
While graph-based retrieval remains valuable for highly complex tasks such as autonomous research and deep research scenarios, its advantages in many multi-hop question-answering scenarios are diminished in agentic RAG systems, where iterative multi-turn querying is already a native capability. 

\begin{figure*}[ht]
\centering
\includegraphics[width=\textwidth]{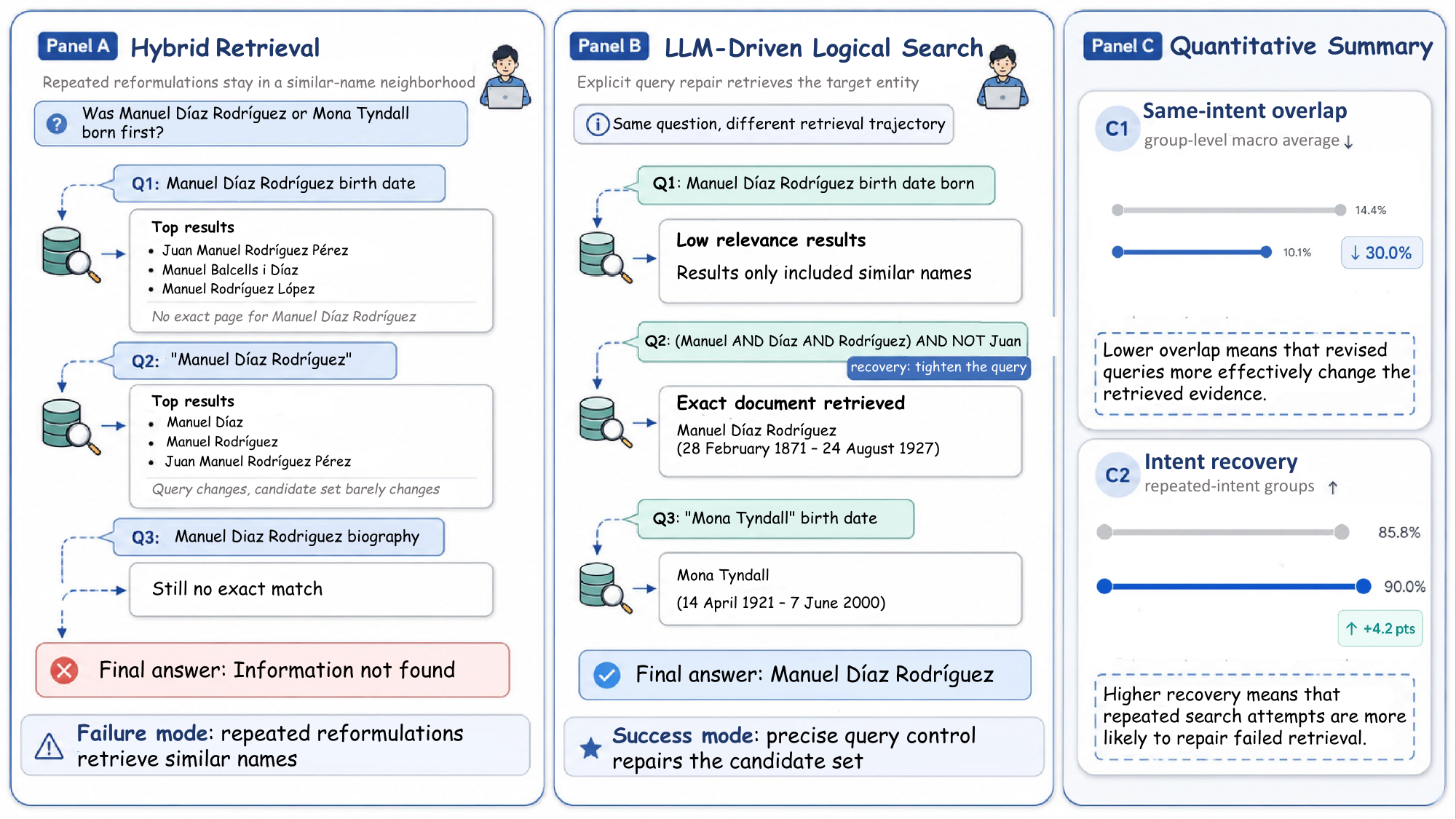}
\caption{
Motivating example and trajectory statistics for LLM-driven logical retrieval.
Agentic Hybrid repeatedly retrieves similar-name distractors, while \textsc{LogicalRAG} repairs the search with explicit lexical constraints.
The right panel reports same-intent result overlap and intent recovery across trajectories.
}
\label{fig:trajectory_motivation}
\vspace{-3mm}
\end{figure*}

Embedding-based retrieval methods were primarily developed to retrieve information relevant to natural-language queries by computing similarity within a dense semantic embedding space \citep{karpukhin-etal-2020-dense}.
Although embeddings are highly effective at capturing abstract semantic relationships, they may overlook critical lexical details, such as proper names or rare entities appearing in queries.
To mitigate this issue, embedding-based retrieval is often augmented with sparse retrieval techniques, such as TF-IDF or other vector space models \citep{arivazhagan-etal-2023-hybrid,wang-etal-2024-searching}.
However, these methods still face a fundamental challenge in multi-turn agentic RAG settings: the retrieved results are often relatively insensitive to slightly refined queries. Because refined queries tend to remain semantically similar to previous ones, the retrieved documents frequently exhibit substantial overlap with earlier retrieval results (Figure~\ref{fig:trajectory_motivation}).
This behavior not only reduces retrieval efficiency by repeatedly returning redundant content, but can also make it difficult for LLMs to repair failed searches through subsequent query refinement.

These observations motivate us to rethink the design of agentic RAG systems. 
We argue that greater control over the retrieval process should be delegated to the LLM during iterative multi-turn interactions.
This paradigm is enabled by the rapidly improving ability of LLMs to formulate structured and precise queries that explicitly capture their retrieval intent. 
Correspondingly, retrieval backends should faithfully execute the search actions issued by the LLM, while exposing a high-control interface that lets the model broaden recall, tighten precision, exclude distractors, and adjust matching granularity.
When the requested information is unavailable, such an interface should also make retrieval failure more explicit, rather than forcing approximate semantic matches that may encourage unsupported answers.


Guided by these design principles for agentic RAG systems, we enable LLMs to construct complex queries using logical operators and compositional constraints (Section~\ref{sec:method}). 
To support such logical query expressions, the underlying retrieval backend is implemented using an inverted-index-based architecture. 
Compared with embedding-based or graph-based retrieval backends, inverted-index systems are substantially easier to build and maintain. 
More importantly, they can faithfully execute explicit lexical constraints and provide a transparent feedback signal for subsequent query refinement.
We found that combining logical query expressions with inverted-index retrieval achieved performance comparable to a strong agentic hybrid baseline while substantially reducing construction and serving cost.
Furthermore, our experiments demonstrate that grounding the retrieval process in explicit logical queries improves abstention behavior and reduces unsupported answers when evidence is unavailable.


Our contributions are threefold:
\vspace{-3mm}
\begin{itemize}
\setlength{\itemsep}{0pt}
\setlength{\parsep}{0pt}
\setlength{\parskip}{0pt}
    \item We propose a design paradigm for agentic RAG centered on LLM-driven retrieval control. This paradigm frames the retrieval process as an interface-control problem. It emphasizes intent-faithful execution and fine-grained control over the search space, enabling agents to deliberately broaden, constrain, and repair retrieval across multi-turn reasoning.
    \item We introduce \textsc{LogicalRAG}, an agentic RAG framework that executes LLM-formulated logical expressions via a lightweight inverted-index backend. It achieves performance comparable to complex hybrid systems while substantially reducing construction and serving cost, with fewer unsupported answers when evidence is unavailable.
    \item We identify an emerging capability threshold for LLM-driven logical retrieval. Model scaling experiments show that logical retrieval lags behind hybrid retrieval with weaker agents, but reaches parity with the strongest agent we evaluate. This suggests that high-control retrieval interfaces are becoming increasingly viable for modern LLM agents.
\end{itemize}


\section{Related Work}
\vspace{-1mm}

\subsection{Dense and Hybrid Retrieval in RAG}

A widely adopted method is dense retrieval, which maps queries and documents into a continuous embedding space to measure semantic similarity. 
To further improve retrieval recall and precision, dense retrieval is often combined with sparse lexical matching methods, giving rise to hybrid retrieval approaches that jointly capture semantic and lexical relevance \citep{karpukhin-etal-2020-dense, gao2023precise, arivazhagan-etal-2023-hybrid}. 
Although hybrid retrieval has become a popular choice for retrieval backends \cite{wang-etal-2024-searching}, its behavior can be difficult for LLMs to anticipate or control during multi-turn interactions. 
In particular, refined conversational queries often remain close to earlier queries in the shared embedding space, leading to substantial overlap in retrieved results. Such redundancy may confuse the LLM and make it more difficult to formulate effective follow-up queries.

\subsection{Graph- and Tree-based Retrieval}

Another line of research introduces explicit structure into the retrieval space. Graph-based systems such as \textsc{GraphRAG} and \textsc{HippoRAG} organize documents around entities and relations, while tree- or page-based methods structure evidence hierarchically to support navigation and aggregation \citep{edge2024local, gutierrez2024hipporag, zhang2025pageindex}. 
Such structured representations can facilitate multi-hop reasoning and global information synthesis. 
However, constructing and maintaining graph- or tree-based retrieval structures is often time-consuming and labor-intensive~\citep{zhuang2025linearrag}. These approaches are further complicated by persistent engineering challenges, including entity disambiguation, cascading errors in relation extraction, and the need to continuously update graph structures as the underlying corpus evolves.

\subsection{Agentic Workflows}

Recent RAG systems have shifted from single-step retrieval toward agentic workflows, in which the LLM iteratively plans, retrieves evidence, observes intermediate results, and revises subsequent actions. Prior work such as ReAct \citep{yao2022react} and IRCoT \citep{trivedi2023interleaving} demonstrates that interleaving reasoning and retrieval enables models to recover from incomplete evidence and better decompose multi-hop questions. 
This shift also changes the role of the retriever: beyond returning relevant documents, it must provide feedback signals that the model can interpret and act upon during reasoning~\citep{li2025agenticragdeepreasoning}. 
Inspired by agent tools such as \texttt{grep}-style search in code environments and structure-aware document navigation \citep{anthropic_tools_2024, zhang2025pageindex}, \textsc{LogicalRAG} maintains a simple yet reliable retrieval backend while exposing logical expressions and compositional constraints as an LLM-driven action space.

\section{Method} \label{sec:method}
\vspace{-1mm}

We propose \textsc{LogicalRAG}, an agentic RAG system built around an LLM-driven logical search interface.
Bypassing architectural overhauls, \textsc{LogicalRAG} exposes a retrieval action space through which the model can express and revise precise search intent.
We first define this interface-level principle, then describe its realization with Boolean query logic, adjustable matching granularity, and BM25 ranking~\citep{robertson2009probabilistic}.

\subsection{LLM-driven Retrieval}
\label{subsec:llm_driven}
\vspace{-1mm}

We formalize LLM-driven retrieval as an interface-level design principle for agentic RAG.
The language model is responsible for deciding what evidence to seek, while the retriever is responsible for executing the model's stated search intent.
Beyond standalone matching strength, this view prioritizes whether the interface lets the agent deliberately change the search space as its reasoning evolves.

Two properties are central to this interface.
First, retrieval should be \emph{intent-faithful}: when the model specifies an entity, phrase, term combination, or exclusion, the retriever should reflect that constraint directly in the candidate set.
Second, the interface should provide sufficient \emph{control} for query repair.
As the model inspects intermediate evidence, it should be able to broaden recall, tighten precision, remove distractors, or adjust matching granularity without relying on the backend to infer these changes implicitly.

This formulation separates the role of the retrieval interface from the complexity of the retrieval backend.
A strong backend may improve standalone retrieval quality, but an agentic system also depends on whether the model can deliberately change what the retriever returns.
\textsc{LogicalRAG} instantiates this principle with a lightweight lexical interface that exposes logical constraints and adjustable matching granularity while using BM25 only to rank matched candidates.

\subsection{\textsc{LogicalRAG}: Controllable and Faithful Keyword Search}
\label{subsec:logicalrag}
\vspace{-1mm}

\textsc{LogicalRAG} realizes LLM-driven search with a lightweight logical retrieval interface.
The agent issues structured lexical queries, while the retriever executes these queries over a standard inverted index and ranks matched passages with BM25.
This design keeps the backend simple and efficient, while giving the LLM direct control over how its search intent is translated into corpus access.

The interface supports two kinds of search control.
First, it supports Boolean query logic, allowing the agent to express required terms, alternatives, and exclusions.
For example, the agent can require an entity to co-occur with a relation phrase, expand recall with aliases, or remove a known distractor.
Second, it supports adjustable matching granularity.
The agent can begin with broad keyword matching for exploration, and later switch to exact phrase or entity-level matching when a specific surface form becomes important.
This is especially useful in multi-hop questions, where early retrieval may identify an intermediate entity and later retrieval must fetch evidence about that entity precisely.

Concretely, we instantiate the interface with a Lucene-style query-string syntax over the \texttt{title} and \texttt{content} fields.
The agent can combine terms with \texttt{AND}, \texttt{OR}, and \texttt{NOT}, group constraints with parentheses, use quoted phrases for exact matching, and optionally target fields such as \texttt{title:"entity name"}.

Given a logical query, the inverted index first determines the candidate documents that satisfy the lexical constraints.
BM25 is then used to rank documents within this candidate set.
Thus, the logical query controls \emph{what can be retrieved}, while BM25 controls \emph{which matched documents are ranked highest}.
This separation makes retrieval behavior more intent-faithful than latent semantic matching: when the agent changes a constraint or matching granularity, the eligible candidate set changes accordingly.

\textsc{LogicalRAG} does not introduce a new scoring function or retrieval architecture. Instead, it provides a lightweight interface tailored for LLM-driven search, combining lexical matching for faithful query execution, logical operators for recovering from failed searches, and BM25 ranking for evidence selection among matched documents. This design enables the LLM to iteratively refine retrieval across interaction turns while avoiding the cost of online query embedding generation, corpus-scale vector indexing, and graph construction.

\section{Experimental Setup}
\label{sec:experimental_setup}
\vspace{-1mm}

To assess its effectiveness–efficiency trade-off in multi-hop QA, we compare \textsc{LogicalRAG} against no-retrieval, hybrid, and graph-based agentic baselines. Our analysis spans answer accuracy, trajectory-level query repair, system efficiency, model scaling, and robustness.

\subsection{Datasets and Scales}
\label{subsec:datasets}

We conducted evaluations under two corpus-scale settings.

\paragraph{Medium-scale corpora.}
Following the dataset construction method of HippoRAG~\citep{gutierrez2024hipporag}, we evaluated on HotpotQA~\citep{yang2018hotpotqa}, 2WikiMultiHopQA~\citep{ho2020constructing}, and MuSiQue~\citep{trivedi2022musique}.
For each dataset, we used 1{,}000 questions and the corresponding dataset-specific Wikipedia subset, containing approximately 9K, 6K, and 11K passages for HotpotQA, 2WikiMultiHopQA, and MuSiQue, respectively.
This setting allows for a fair comparison with graph-based and agentic RAG baselines under controlled corpus sizes.

\paragraph{Large-scale corpus.}
We further evaluated \textsc{LogicalRAG} on the KILT-scale Wikipedia corpus~\citep{petroni-etal-2021-kilt}.
Unlike the medium-scale setting, queries are matched against a substantially larger open-domain knowledge base.
This setting places greater demands on indexing efficiency, serving latency, and the ability of each retrieval interface to support multi-step agentic search.
Graph-based systems are not evaluated at this scale, as their released pipelines require corpus-level graph construction, which is prohibitively expensive in terms of both time and computational resources.

\subsection{Baseline Methods and Systems}
\label{subsec:baselines}

We compared \textsc{LogicalRAG} with four baselines.
\textbf{Direct Answer} produces answers without retrieval and serves as a no-evidence reference point.
\textbf{Agentic Hybrid} is our main controlled baseline: it uses the same multi-step search procedure as \textsc{LogicalRAG}, but replaces logical lexical search with hybrid sparse-dense retrieval merged by reciprocal rank fusion.
\textbf{HippoRAG2}~\citep{gutierrez2025rag} represents graph-based RAG and is evaluated only in the medium-scale setting, where graph construction is feasible.
\textbf{MA-RAG}~\citep{nguyen2025ma} is an embedding-based agentic RAG system with a more engineered multi-agent harness, including sub-agents and parallel retrieval calls, and is evaluated as an end-to-end baseline.

For the controlled comparison between \textsc{LogicalRAG} and Agentic Hybrid, both systems use the same lightweight search procedure.
Given a question, the model first generates a tentative evidence-seeking plan, then iteratively observes retrieved passages and either issues another retrieval action or produces a final answer.
The two systems differ only in retrieval action space: \textsc{LogicalRAG} issues structured lexical queries with logical constraints and adjustable matching granularity, while Agentic Hybrid issues natural-language queries to a hybrid retriever.
We intentionally kept the search procedure simple by appending previous searches and retrieved passages to the running context, without specialized memory compression, reranking, or additional controllers.
Complete prompts, retrieval limits, RRF parameters, and generation settings are provided in Appendix~\ref{app:implementation}.

\subsection{Efficiency Evaluation}
\label{subsec:cost_protocol}

We measured offline construction cost and online retrieval efficiency on the KILT-scale corpus.
Offline cost includes all preprocessing before serving: inverted-index construction for both systems, plus corpus-wide embedding generation and FAISS indexing for Agentic Hybrid~\citep{douze2025faiss}.
All indexing and serving measurements were conducted on the same local server; hardware and serving details are provided in Appendix~\ref{app:implementation}.

For online efficiency, we replayed retrieval queries from KILT-scale trajectories after warming up the retrieval services.
We reported throughput and per-request wall-clock latency under different client-side concurrency levels.
\textsc{LogicalRAG} was measured as a full OpenSearch retrieval call~\citep{opensearch_github}.
Agentic Hybrid was measured as a full hybrid retrieval call, including sparse retrieval, query embedding, dense retrieval, document fetching, and rank fusion.
Mean and P95 latencies were computed over individual request wall-clock times at each concurrency level.
Additional phase-level measurements are provided in Appendix~\ref{app:implementation}.

\subsection{Evaluation Metrics}
\label{subsec:metrics}

\paragraph{Answer accuracy.}
We reported exact match (EM), word-level F1, and LLM-as-a-Judge accuracy.
EM and F1 evaluate surface-form correctness, while LLM-as-a-Judge additionally captures semantically equivalent answers that may differ lexically.

\paragraph{Trajectory-level query repair.}
\vspace{-1mm}
We grouped retrieval actions by their underlying search intent and report same-intent result overlap and intent recovery rate.
These metrics measure whether query revisions change the retrieved evidence and recover from earlier failed searches.
Additional details are provided in Appendix~\ref{app:trajectory_metrics}.

\paragraph{System efficiency.}
\vspace{-1mm}
We reported offline index construction cost, online throughput, and per-query retrieval latency, following the efficiency setup described in Section~\ref{subsec:cost_protocol}.

\paragraph{Answer-unavailable robustness.}
\vspace{-1mm}
We removed gold supporting passages to create answer-unavailable examples and report refusal, hallucination, and gold leak rates.
Hallucination denotes an incorrect, non-refusal answer unsupported by the remaining corpus, while gold leak denotes a correct answer despite evidence removal.

\paragraph{Model scaling.}
\vspace{-1mm}
We evaluated Agentic Hybrid and \textsc{LogicalRAG} using multiple Qwen3.5 models~\citep{qwen35blog} on the KILT-scale corpus.
This experiment investigates how retrieval performance scales with agent capability and examines whether stronger agents can more effectively exploit the high-control action space enabled by LLM-driven logical search.

\begin{table*}[t!]
\centering
\small
\setlength{\tabcolsep}{7pt}
\renewcommand{\arraystretch}{1.08}
\definecolor{sectionblue}{RGB}{226,244,250}
\definecolor{lightgrayrow}{RGB}{245,245,245}

\begin{tabular}{lcccccccccc}
\toprule
\multirow{2}{*}{Method}
& \multicolumn{3}{c}{2WikiMultiHopQA}
& \multicolumn{3}{c}{HotpotQA}
& \multicolumn{3}{c}{MuSiQue}
& \multirow{2}{*}{Avg. Judge} \\
\cmidrule(lr){2-4}
\cmidrule(lr){5-7}
\cmidrule(lr){8-10}
& EM & F1 & Judge
& EM & F1 & Judge
& EM & F1 & Judge
& \\
\midrule

\rowcolor{lightgrayrow}
Direct Answer
& 0.383 & 0.508 & 0.596
& 0.467 & 0.546 & 0.597
& 0.200 & 0.324 & 0.340
& 0.511 \\
\midrule

\rowcolor{sectionblue}
\multicolumn{11}{c}{\textit{Medium-scale corpus}} \\
\midrule
HippoRAG2
& 0.575 & 0.655 & 0.732
& 0.560 & 0.689 & 0.802
& 0.366 & 0.489 & 0.515
& 0.683 \\
MA-RAG
& 0.663 & 0.766 & 0.860
& 0.536 & 0.669 & 0.767
& 0.335 & 0.485 & 0.545
& 0.724 \\
Agentic Hybrid
& 0.706 & 0.785 & 0.867
& \textbf{0.623} & \textbf{0.762} & \textbf{0.893}
& \textbf{0.461} & \textbf{0.573} & \textbf{0.662}
& \textbf{0.807} \\
\textsc{LogicalRAG}
& \textbf{0.708} & \textbf{0.789} & \textbf{0.874}
& 0.622 & 0.761 & 0.884
& 0.410 & 0.520 & 0.595
& 0.784 \\
\midrule

\rowcolor{sectionblue}
\multicolumn{11}{c}{\textit{Large-scale corpus (KILT)}} \\
\midrule
HippoRAG2
& \multicolumn{3}{c}{N/A}
& \multicolumn{3}{c}{N/A}
& \multicolumn{3}{c}{N/A}
& N/A \\
MA-RAG
& 0.429 & 0.501 & 0.543
& 0.337 & 0.457 & 0.523
& 0.137 & 0.252 & 0.273
& 0.446 \\
Agentic Hybrid
& \textbf{0.713} & \textbf{0.797} & 0.882
& \textbf{0.534} & \textbf{0.668} & \textbf{0.794}
& \textbf{0.302} & 0.420 & 0.471
& 0.716 \\
\textsc{LogicalRAG}
& 0.701 & 0.787 & \textbf{0.886}
& 0.524 & 0.664 & 0.784
& \textbf{0.302} & \textbf{0.422} & \textbf{0.481}
& \textbf{0.717} \\
\bottomrule
\end{tabular}
\caption{
Main experimental results on medium-scale and KILT-scale corpora.
``Judge'' denotes LLM-as-a-Judge accuracy.
Avg. Judge is the average Judge score across the three datasets within each corpus setting.
}
\label{tab:main_results}
\vspace{-3mm}
\end{table*}

\section{Results}
\label{sec:results}

We evaluated \textsc{LogicalRAG} from five perspectives:
(i) answer accuracy under medium-scale and KILT-scale corpora,
(ii) trajectory-level query refinement,
(iii) system efficiency,
(iv) how performance changes with agent capability, and
(v) robustness when the answer is absent from the knowledge base.
Overall, results showed that a lightweight lexical backend could be competitive with more complex retrieval systems when the interface faithfully executes the LLM's search intent and provides fine-grained retrieval control.

\subsection{Main Results: Matching Hybrid Retrieval at Scale}
\label{subsec:main_results}

\begin{table*}[t]
\centering
\small
\setlength{\tabcolsep}{8pt}
\begin{tabular}{llrr}
\toprule
Backend & Main preprocessing & Construction Cost & Relative \\
\midrule
\textsc{LogicalRAG}
& Inverted index
& $1.27$ h
& $1.0\times$ \\
Agentic Hybrid
& Inverted index + Embedding + FAISS
& $52.02$ h
& $41.0\times$ \\
Graph-based
& LLM graph construction
& $\sim$37B input + 18B output tokens
& Outside budget \\
\bottomrule
\end{tabular}
\caption{
Offline construction cost on the KILT-scale corpus. Graph-based construction cost is measured in terms of LLM preprocessing tokens.
}
\label{tab:indexing_cost}
\vspace{-2mm}
\end{table*}

\begin{table}[t]
\centering
\small
\setlength{\tabcolsep}{6pt}
\renewcommand{\arraystretch}{1.06}
\begin{tabular}{lrrr}
\toprule
Backend & QPS $\uparrow$ & Mean ms $\downarrow$ & P95 ms $\downarrow$ \\
\midrule
\multicolumn{4}{l}{\textit{Concurrency = 1}} \\
\quad \textsc{LogicalRAG} & 11.8 & 84.6  & 232.9 \\
\quad Hybrid              & 5.4  & 184.5 & 279.1 \\
\midrule
\multicolumn{4}{l}{\textit{Concurrency = 16}} \\
\quad \textsc{LogicalRAG} & 152.5 & 74.9  & 234.1 \\
\quad Hybrid              & 66.6  & 230.5 & 294.2 \\
\bottomrule
\end{tabular}
\caption{
Online retrieval efficiency on KILT-scale trajectory queries.
Latencies are per-request wall-clock times.
}
\label{tab:latency}
\vspace{-3mm}
\end{table}

Table~\ref{tab:main_results} reports the main results on medium-scale and KILT-scale corpora.
Across both settings, \textsc{LogicalRAG} was competitive with Agentic Hybrid, despite using only lexical matching, logical query control, and BM25 ranking.
On the medium-scale corpora, \textsc{LogicalRAG} slightly outperformed Agentic Hybrid on 2WikiMultiHopQA, while Agentic Hybrid remained stronger on HotpotQA and MuSiQue.
On the KILT-scale corpus, the two systems reached near parity: the average LLM-as-a-Judge accuracy is 0.717 for \textsc{LogicalRAG} and 0.716 for Agentic Hybrid.

This result is notable because \textsc{LogicalRAG} does not add dense retrieval to the backend.
Instead, it relies on the LLM to drive search through explicit lexical constraints.
The trajectory analysis in Figure~\ref{fig:trajectory_motivation} helps explain why this simple interface can be effective.
When the agent repeatedly searched for the same information need, \textsc{LogicalRAG} retrieved result sets with lower overlap than Agentic Hybrid and achieved higher recovery from failed searches.
This suggests that logical query control helps the agent repair retrieval failures by changing the candidate set more directly, rather than repeatedly retrieving similar evidence.

These results highlight a key observation: in agentic RAG, strong end-to-end performance does not necessarily require increasingly complex retrievers.
When the retrieval interface can faithfully execute the LLM’s explicit search intent while providing sufficient control for iterative refinement across turns, a lightweight lexical backend can achieve performance comparable to that of a substantially more expensive hybrid retriever.
Having established comparable answer accuracy, we next examined the system cost required to achieve this performance.

\subsection{Efficiency: Matching Accuracy with Lower Retrieval Cost}
\label{subsec:efficiency}

Table~\ref{tab:indexing_cost} shows that the comparable accuracy in Table~\ref{tab:main_results} comes with substantially different construction costs.
\textsc{LogicalRAG} only builds an inverted index, whereas Agentic Hybrid additionally requires corpus-wide embedding generation and FAISS indexing, resulting in about \(41\times\) higher offline construction time.
Graph-based retrieval is even less aligned with large-scale deployment, requiring LLM-based entity and relation extraction over the corpus, which amounts to about 37B input tokens and 18B output tokens in our estimate.

The efficiency gain is not attributable to a shorter agent loop, as the two controlled systems use a comparable number of search turns: $3.77$ per question for \textsc{LogicalRAG} versus $3.55$ for Agentic Hybrid.
Instead, the improvement arises from the lighter-weight retrieval backend during online serving.
At a concurrency level of $16$, \textsc{LogicalRAG} achieved $152.5$ QPS with a mean latency of $74.9$ ms, whereas Agentic Hybrid reached $66.6$ QPS with a mean latency of $230.5$ ms (Table~\ref{tab:latency}).
This corresponds to $2.3 \times$ higher throughput and $3.1\times$ lower mean latency for \textsc{LogicalRAG}.

Together, these results showed that \textsc{LogicalRAG} reached near-parity in answer accuracy while avoiding both corpus-wide embedding construction and online query embedding.
By executing the LLM's logical query directly over an inverted index, it retains ranked top-\(k\) retrieval with BM25 at substantially lower construction and serving cost.
Additional serving details and phase-level measurements are provided in Appendix~\ref{app:implementation}.

\subsection{Model Scaling: Stronger Agents Make LLM-Driven Search Viable}
\label{subsec:model_scaling}

A central question for LLM-driven search is why such a lightweight retrieval design becomes competitive.
Our hypothesis is that logical search becomes more effective as the agent becomes better at decomposing questions, inspecting evidence, and issuing precise query revisions.
We therefore evaluated Agentic Hybrid and \textsc{LogicalRAG} on the KILT-scale corpus with multiple Qwen3.5 models.

\begin{figure}[ht]
\centering
\includegraphics[width=\linewidth]{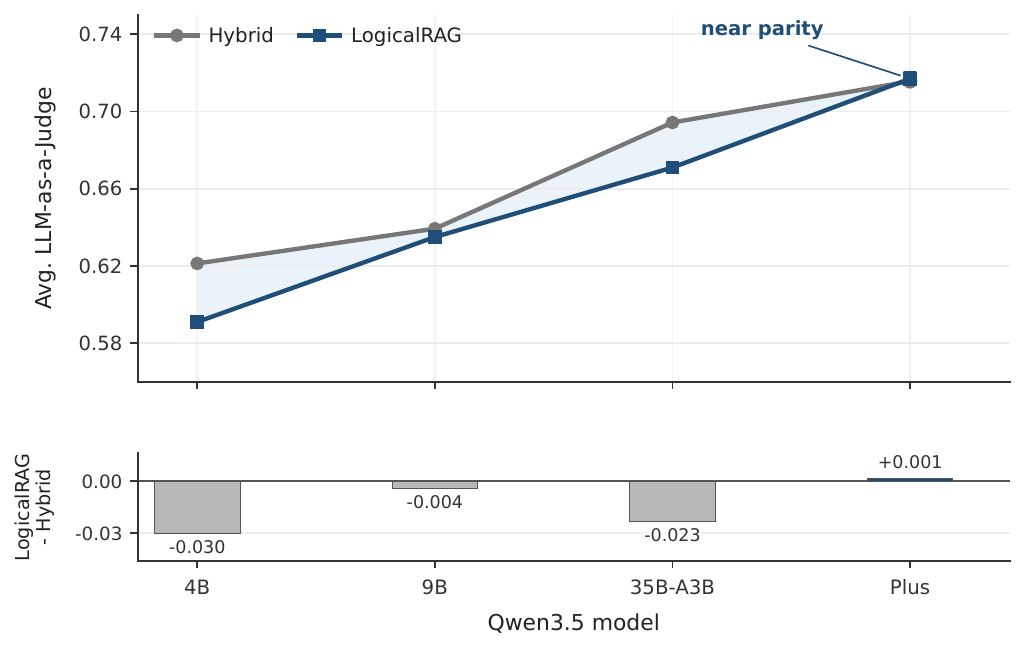}
\caption{
Model scaling on the KILT-scale corpus. As model capability increases, \textsc{LogicalRAG} progressively closes the gap with Agentic Hybrid, reaching near parity on Qwen3.5-Plus.
}
\label{fig:model_scaling}
\vspace{-3mm}
\end{figure}

Figure~\ref{fig:model_scaling} reports the average LLM-as-a-Judge accuracy across 2WikiMultiHopQA, HotpotQA, and MuSiQue.
The overall trend supports this hypothesis.
With Qwen3.5-4B, \textsc{LogicalRAG} trailed Agentic Hybrid by about three points, indicating that smaller agents still benefit from the additional semantic smoothing of hybrid retrieval.
As the agent became stronger, \textsc{LogicalRAG} improved rapidly and the gap to Agentic Hybrid largely closed.
With Qwen3.5-Plus, the two methods reached near parity.
The results suggest that LLM-driven search becomes more viable when the agent is capable enough to operate a high-control retrieval interface, turning reasoning intent into explicit lexical constraints and query repairs.

\subsection{Answer-Unavailable Questions: Reducing Hallucination}
\label{subsec:hallucination}

\begin{table*}[t]
\centering
\small
\setlength{\tabcolsep}{12pt}
\begin{tabular}{llccc}
\toprule
Dataset & \multicolumn{1}{c}{Method} & Refusal $\uparrow$ & Hallucination $\downarrow$ & Gold Leak \\
\midrule
\multirow{2}{*}{2Wiki}
& Hybrid & $0.913$ & $0.050$ & $0.038$ \\
& \textsc{LogicalRAG} & ${\bf 0.943}$ & ${\bf 0.028}$ & $0.030$ \\
\midrule
\multirow{2}{*}{HotpotQA}
& Hybrid & $0.660$ & $0.160$ & $0.180$ \\
& \textsc{LogicalRAG} & ${\bf 0.733}$ & ${\bf 0.110}$ & $0.158$ \\
\midrule
\multirow{2}{*}{MuSiQue}
& Hybrid & $0.728$ & $0.175$ & $0.098$ \\
& \textsc{LogicalRAG} & ${\bf 0.808}$ & ${\bf 0.110}$ & $0.083$ \\
\midrule
\multirow{2}{*}{Avg.}
& Hybrid & $0.767$ & $0.128$ & $0.105$ \\
& \textsc{LogicalRAG} & ${\bf 0.828}$ & ${\bf 0.083}$ & $0.090$ \\
\bottomrule
\end{tabular}
\caption{
Answer-unavailable evaluation after removing gold supporting passages.
Hallucination denotes an incorrect non-refusal answer unsupported by the remaining corpus.
}
\label{tab:hallucination}
\vspace{-3mm}
\end{table*}

We evaluated robustness when the corpus lacks sufficient evidence to answer the question.
For each dataset, we constructed 400 answer-unavailable examples by removing gold supporting passages.
The desired behavior is to refuse rather than produce an unsupported answer.
We report refusal, hallucination, and gold leak rates, where gold leak denotes a correct answer despite evidence removal, and hallucination denotes an incorrect non-refusal answer unsupported by the remaining corpus.

Table~\ref{tab:hallucination} shows that \textsc{LogicalRAG} improved abstention behavior across all three datasets.
On average, its refusal rate increased from $0.767$ to $0.828$, while the hallucination rate decreased from $0.128$ to $0.083$.

We attribute this improvement to the legibility of failed logical search.
Repeated failures under explicit lexical constraints provide a clearer signal that required evidence may be absent, whereas Agentic Hybrid may still return semantically related but unsupported passages.
Thus, LLM-driven logical search helps the agent distinguish missing evidence from merely incomplete retrieval, leading to more appropriate abstention.

\section{Discussion}
\label{sec:discussion}

\paragraph{Logical search and \texttt{grep}-style access.}
\vspace{-1mm}
Recent agent systems have explored direct text access through \texttt{grep}-style tools~\citep{subramanian2025keyword,wang2026greprag,li2026semanticsimilarityrethinkingretrieval}.
\textsc{LogicalRAG} shares the same motivation of giving the agent explicit lexical control, but differs in the retrieval substrate.
A literal \texttt{grep} interface is effective for exact string matching, yet it provides limited ranking and can be inefficient for large-scale top-\(k\) evidence retrieval.
\textsc{LogicalRAG} implements lexical control with a standard inverted-index engine, supporting Boolean logic, phrase matching, and BM25 ranking while retaining efficient retrieval over large corpora.
Thus, \texttt{grep}-style search can be viewed as a useful special case of direct lexical access, whereas \textsc{LogicalRAG} targets ranked, large-scale RAG over document collections.

\paragraph{Why not combine dense and logical tools?}
\vspace{-1mm}
A natural extension is to expose both dense retrieval and logical search to the agent~\citep{du2026rag}.
Such a system may improve accuracy when relevant evidence is difficult to express lexically.
We did not study this combined setting as our main system because it answers a different question.
Our goal was to isolate how far LLM-driven search can go with a lightweight lexical backend under a controlled agentic loop.
Adding dense retrieval would reintroduce corpus-wide embedding construction and online query encoding, reducing the efficiency advantage that motivates our design.
Hybrid tool use is therefore a promising extension, but not the focus of this work.

\paragraph{Interface ablations.}
Additional ablations in Appendix~\ref{app:ablation} suggest that \textsc{LogicalRAG} is not highly sensitive to prompt engineering.
Removing Boolean operators moderately reduces performance, while replacing the full tool description with a minimal syntax-only description causes little degradation.
These results support our interpretation that an intent-faithful lexical search interface drives the main performance benefit, independent of prompt engineering.

\section{Conclusion}
In this study, we revisited the design of agentic RAG and challenged the prevailing trend of building increasingly complex retrieval backends. We proposed a shift in perspective toward a ``heavy brain, precise hands'' paradigm, where the complex reasoning and query planning are centralized within the LLM (the ``heavy brain''), while the retrieval backend is simplified into a lightweight, deterministic execution layer (the ``precise hands''). 
Implementing this philosophy, we introduced \textsc{LogicalRAG}, a framework that empowers LLMs to formulate precise retrieval intents using compositional logical expressions, executed over a lightweight inverted-index backend. 
Our evaluations demonstrate that this framework matches the performance of dense, hybrid, and graph-based baselines, while achieving significant reductions in downstream hallucinations and system overhead.
More broadly, our findings suggest that agentic RAG may benefit from a ``heavy brain, precise hands'' paradigm, in which increasingly capable LLMs perform sophisticated retrieval planning and reasoning, while retrieval systems serve as faithful executors of the model’s structured intent.

\section*{Limitations}
Although \textsc{LogicalRAG} demonstrates strong performance across a wide range of tasks, several limitations remain. 
First, our framework depends on the ability of LLMs to formulate accurate logical queries. When the model fails to identify the appropriate constraints, entities, or query structure, retrieval quality may degrade correspondingly. 
Second, while logical retrieval provides strong precision and interpretability, embedding-based retrieval may still offer advantages in scenarios requiring highly abstract semantic matching or fuzzy associations that are difficult to express explicitly through logical constraints. 
To alleviate this, our retrieval backend also supports natural language access, although such retrieval is generally less accurate than embedding-based approaches.
In addition, our current implementation primarily focuses on textual retrieval settings, and extending the framework to multimodal or continuously evolving knowledge environments remains an important direction for future work. 
Nevertheless, we believe these limitations do not diminish the broader promise of the proposed paradigm for building simpler, more controllable, and more reliable agentic RAG systems.

\bibliography{custom}

@inproceedings{NEURIPS2020_6b493230,
 author = {Lewis, Patrick and Perez, Ethan and Piktus, Aleksandra and Petroni, Fabio and Karpukhin, Vladimir and Goyal, Naman and K\"{u}ttler, Heinrich and Lewis, Mike and Yih, Wen-tau and Rockt\"{a}schel, Tim and Riedel, Sebastian and Kiela, Douwe},
 booktitle = {Advances in Neural Information Processing Systems},
 editor = {H. Larochelle and M. Ranzato and R. Hadsell and M.F. Balcan and H. Lin},
 pages = {9459--9474},
 publisher = {Curran Associates, Inc.},
 title = {Retrieval-Augmented Generation for Knowledge-Intensive NLP Tasks},
 url = {https://proceedings.neurips.cc/paper_files/paper/2020/file/6b493230205f780e1bc26945df7481e5-Paper.pdf},
 volume = {33},
 year = {2020}
}

@inproceedings{izacard-grave-2021-leveraging,
    title = "Leveraging Passage Retrieval with Generative Models for Open Domain Question Answering",
    author = "Izacard, Gautier  and
      Grave, Edouard",
    editor = "Merlo, Paola  and
      Tiedemann, Jorg  and
      Tsarfaty, Reut",
    booktitle = "Proceedings of the 16th Conference of the European Chapter of the Association for Computational Linguistics: Main Volume",
    month = apr,
    year = "2021",
    address = "Online",
    publisher = "Association for Computational Linguistics",
    url = "https://aclanthology.org/2021.eacl-main.74/",
    doi = "10.18653/v1/2021.eacl-main.74",
    pages = "874--880"
}

@inproceedings{wang-etal-2024-searching,
    title = "Searching for Best Practices in Retrieval-Augmented Generation",
    author = "Wang, Xiaohua  and
      Wang, Zhenghua  and
      Gao, Xuan  and
      Zhang, Feiran  and
      Wu, Yixin  and
      Xu, Zhibo  and
      Shi, Tianyuan  and
      Wang, Zhengyuan  and
      Li, Shizheng  and
      Qian, Qi  and
      Yin, Ruicheng  and
      Lv, Changze  and
      Zheng, Xiaoqing  and
      Huang, Xuanjing",
    editor = "Al-Onaizan, Yaser  and
      Bansal, Mohit  and
      Chen, Yun-Nung",
    booktitle = "Proceedings of the 2024 Conference on Empirical Methods in Natural Language Processing",
    month = nov,
    year = "2024",
    address = "Miami, Florida, USA",
    publisher = "Association for Computational Linguistics",
    url = "https://aclanthology.org/2024.emnlp-main.981/",
    doi = "10.18653/v1/2024.emnlp-main.981",
    pages = "17716--17736"
}

@inproceedings{karpukhin-etal-2020-dense,
    title = "Dense Passage Retrieval for Open-Domain Question Answering",
    author = "Karpukhin, Vladimir  and
      Oguz, Barlas  and
      Min, Sewon  and
      Lewis, Patrick  and
      Wu, Ledell  and
      Edunov, Sergey  and
      Chen, Danqi  and
      Yih, Wen-tau",
    editor = "Webber, Bonnie  and
      Cohn, Trevor  and
      He, Yulan  and
      Liu, Yang",
    booktitle = "Proceedings of the 2020 Conference on Empirical Methods in Natural Language Processing (EMNLP)",
    month = nov,
    year = "2020",
    address = "Online",
    publisher = "Association for Computational Linguistics",
    url = "https://aclanthology.org/2020.emnlp-main.550/",
    doi = "10.18653/v1/2020.emnlp-main.550",
    pages = "6769--6781"
}

@inproceedings{arivazhagan-etal-2023-hybrid,
    title = "Hybrid Hierarchical Retrieval for Open-Domain Question Answering",
    author = "Arivazhagan, Manoj Ghuhan  and
      Liu, Lan  and
      Qi, Peng  and
      Chen, Xinchi  and
      Wang, William Yang  and
      Huang, Zhiheng",
    editor = "Rogers, Anna  and
      Boyd-Graber, Jordan  and
      Okazaki, Naoaki",
    booktitle = "Findings of the Association for Computational Linguistics: ACL 2023",
    month = jul,
    year = "2023",
    address = "Toronto, Canada",
    publisher = "Association for Computational Linguistics",
    url = "https://aclanthology.org/2023.findings-acl.679/",
    doi = "10.18653/v1/2023.findings-acl.679",
    pages = "10680--10689"
}

@article{edge2024local,
  title={From local to global: A graph rag approach to query-focused summarization},
  author={Edge, Darren and Trinh, Ha and Cheng, Newman and Bradley, Joshua and Chao, Alex and Mody, Apurva and Truitt, Steven and Metropolitansky, Dasha and Ness, Robert Osazuwa and Larson, Jonathan},
  journal={arXiv preprint arXiv:2404.16130},
  year={2024}
}

@inproceedings{yao2022react,
  title={{React}: Synergizing reasoning and acting in language models},
  author={Yao, Shunyu and Zhao, Jeffrey and Yu, Dian and Du, Nan and Shafran, Izhak and Narasimhan, Karthik R and Cao, Yuan},
  booktitle={The eleventh international conference on learning representations},
  year={2022}
}

@inproceedings{trivedi2023interleaving,
  title={Interleaving retrieval with chain-of-thought reasoning for knowledge-intensive multi-step questions},
  author={Trivedi, Harsh and Balasubramanian, Niranjan and Khot, Tushar and Sabharwal, Ashish},
  booktitle={Proceedings of the 61st annual meeting of the association for computational linguistics (volume 1: long papers)},
  pages={10014--10037},
  year={2023}
}

@misc{anthropic_tools_2024,
  author = {{Anthropic}},
  title = {Claude Code: Tools Reference},
  year = {2024},
  url = {https://code.claude.com/docs/en/tools-reference},
  note = {Accessed: 2026-04-21}
}

@article{zhang2025pageindex,
  author = {Mingtian Zhang and Yu Tang and PageIndex Team},
  title = {{PageIndex}: Next-Generation Vectorless, Reasoning-based RAG},
  journal = {PageIndex Blog},
  year = {2025},
  month = {September},
  note = {{https://pageindex.ai/blog/pageindex-intro}},
}

@book{robertson2009probabilistic,
  title={The probabilistic relevance framework: BM25 and beyond},
  author={Robertson, Stephen and Zaragoza, Hugo},
  volume={4},
  year={2009},
  publisher={Now Publishers Inc}
}

@inproceedings{yang2018hotpotqa,
  title={{HotpotQA}: A dataset for diverse, explainable multi-hop question answering},
  author={Yang, Zhilin and Qi, Peng and Zhang, Saizheng and Bengio, Yoshua and Cohen, William and Salakhutdinov, Ruslan and Manning, Christopher D},
  booktitle={Proceedings of the 2018 conference on empirical methods in natural language processing},
  pages={2369--2380},
  year={2018}
}

@article{gutierrez2024hipporag,
  title={{Hipporag}: Neurobiologically inspired long-term memory for large language models},
  author={Guti{\'e}rrez, Bernal J and Shu, Yiheng and Gu, Yu and Yasunaga, Michihiro and Su, Yu},
  journal={Advances in neural information processing systems},
  volume={37},
  pages={59532--59569},
  year={2024}
}

@inproceedings{ho2020constructing,
  title={Constructing a multi-hop qa dataset for comprehensive evaluation of reasoning steps},
  author={Ho, Xanh and Nguyen, Anh-Khoa Duong and Sugawara, Saku and Aizawa, Akiko},
  booktitle={Proceedings of the 28th International Conference on Computational Linguistics},
  pages={6609--6625},
  year={2020}
}

@article{trivedi2022musique,
  title={{MuSiQue}: Multihop Questions via Single-hop Question Composition},
  author={Trivedi, Harsh and Balasubramanian, Niranjan and Khot, Tushar and Sabharwal, Ashish},
  journal={Transactions of the Association for Computational Linguistics},
  volume={10},
  pages={539--554},
  year={2022},
  publisher={MIT Press One Broadway, 12th Floor, Cambridge, Massachusetts 02142, USA~…}
}

@inproceedings{petroni-etal-2021-kilt,
    title = "{KILT}: a Benchmark for Knowledge Intensive Language Tasks",
    author = {Petroni, Fabio  and
      Piktus, Aleksandra  and
      Fan, Angela  and
      Lewis, Patrick  and
      Yazdani, Majid  and
      De Cao, Nicola  and
      Thorne, James  and
      Jernite, Yacine  and
      Karpukhin, Vladimir  and
      Maillard, Jean  and
      Plachouras, Vassilis  and
      Rockt{\"a}schel, Tim  and
      Riedel, Sebastian},
    editor = "Toutanova, Kristina  and
      Rumshisky, Anna  and
      Zettlemoyer, Luke  and
      Hakkani-Tur, Dilek  and
      Beltagy, Iz  and
      Bethard, Steven  and
      Cotterell, Ryan  and
      Chakraborty, Tanmoy  and
      Zhou, Yichao",
    booktitle = "Proceedings of the 2021 Conference of the North American Chapter of the Association for Computational Linguistics: Human Language Technologies",
    month = jun,
    year = "2021",
    address = "Online",
    publisher = "Association for Computational Linguistics",
    url = "https://aclanthology.org/2021.naacl-main.200/",
    doi = "10.18653/v1/2021.naacl-main.200",
    pages = "2523--2544"
}

@article{nguyen2025ma,
  title={{Ma-rag}: Multi-agent retrieval-augmented generation via collaborative chain-of-thought reasoning},
  author={Nguyen, Thang and Chin, Peter and Tai, Yu-Wing},
  journal={arXiv preprint arXiv:2505.20096},
  year={2025}
}

@article{gutierrez2025rag,
  title={From rag to memory: Non-parametric continual learning for large language models},
  author={Guti{\'e}rrez, Bernal Jim{\'e}nez and Shu, Yiheng and Qi, Weijian and Zhou, Sizhe and Su, Yu},
  journal={arXiv preprint arXiv:2502.14802},
  year={2025}
}

@article{douze2025faiss,
  title={The faiss library},
  author={Douze, Matthijs and Guzhva, Alexandr and Deng, Chengqi and Johnson, Jeff and Szilvasy, Gergely and Mazar{\'e}, Pierre-Emmanuel and Lomeli, Maria and Hosseini, Lucas and J{\'e}gou, Herv{\'e}},
  journal={IEEE Transactions on Big Data},
  year={2025},
  publisher={IEEE}
}

@misc{qwen35blog,
    title = {Qwen3.5: Accelerating Productivity with Native Multimodal Agents},
    url = {https://qwen.ai/blog?id=qwen3.5},
    author = {Qwen Team},
    month = {February},
    year = {2026}
}

@article{wang2026greprag,
  title={{GrepRAG}: An Empirical Study and Optimization of Grep-Like Retrieval for Code Completion},
  author={Wang, Baoyi and Wang, Xingliang and Li, Guochang and Zhi, Chen and Han, Junxiao and Zhao, Xinkui and Wang, Nan and Deng, Shuiguang and Yin, Jianwei},
  journal={arXiv preprint arXiv:2601.23254},
  year={2026}
}

@article{subramanian2025keyword,
  title={Keyword search is all you need: Achieving RAG-Level Performance without vector databases using agentic tool use},
  author={Subramanian, Shreyas and Akinfaderin, Adewale and Zhang, Yanyan and Singh, Ishan and Khanuja, Mani and Singh, Sandeep and Tanke, Maira Ladeira},
  journal={arXiv preprint arXiv:2602.23368},
  year={2025}
}

@article{du2026rag,
  title={{A-RAG}: Scaling Agentic Retrieval-Augmented Generation via Hierarchical Retrieval Interfaces},
  author={Du, Mingxuan and Xu, Benfeng and Zhu, Chiwei and Wang, Shaohan and Wang, Pengyu and Wang, Xiaorui and Mao, Zhendong},
  journal={arXiv preprint arXiv:2602.03442},
  year={2026}
}

@inproceedings{ICLR2024_25f7be96,
 author = {Asai, Akari and Wu, Zeqiu and Wang, Yizhong and Sil, Avi and Hajishirzi, Hannaneh },
 booktitle = {International Conference on Learning Representations},
 editor = {B. Kim and Y. Yue and S. Chaudhuri and K. Fragkiadaki and M. Khan and Y. Sun},
 pages = {9112--9141},
 title = {Self-RAG: Learning to Retrieve, Generate, and Critique through Self-Reflection},
 url = {https://proceedings.iclr.cc/paper_files/paper/2024/file/25f7be9694d7b32d5cc670927b8091e1-Paper-Conference.pdf},
 volume = {2024},
 year = {2024}
}

@inproceedings{li2025search,
  title={{Search-o1}: Agentic search-enhanced large reasoning models},
  author={Li, Xiaoxi and Dong, Guanting and Jin, Jiajie and Zhang, Yuyao and Zhou, Yujia and Zhu, Yutao and Zhang, Peitian and Dou, Zhicheng},
  booktitle={Proceedings of the 2025 Conference on Empirical Methods in Natural Language Processing},
  pages={5420--5438},
  year={2025}
}

@inproceedings{zhu2025knowledge,
  title={Knowledge graph-guided retrieval augmented generation},
  author={Zhu, Xiangrong and Xie, Yuexiang and Liu, Yi and Li, Yaliang and Hu, Wei},
  booktitle={Proceedings of the 2025 Conference of the Nations of the Americas Chapter of the Association for Computational Linguistics: Human Language Technologies (Volume 1: Long Papers)},
  pages={8912--8924},
  year={2025}
}

@article{zhuang2025linearrag,
  title={{Linearrag}: Linear graph retrieval augmented generation on large-scale corpora},
  author={Zhuang, Luyao and Chen, Shengyuan and Xiao, Yilin and Zhou, Huachi and Zhang, Yujing and Chen, Hao and Zhang, Qinggang and Huang, Xiao},
  journal={arXiv preprint arXiv:2510.10114},
  year={2025}
}

@misc{opensearch_github,
  title = {{OpenSearch}},
  author = {{OpenSearch Project}},
  year = {2026},
  howpublished = {\url{https://github.com/opensearch-project/OpenSearch}}
}

@inproceedings{gao2023precise,
  title={Precise zero-shot dense retrieval without relevance labels},
  author={Gao, Luyu and Ma, Xueguang and Lin, Jimmy and Callan, Jamie},
  booktitle={Proceedings of the 61st Annual Meeting of the Association for Computational Linguistics (Volume 1: Long Papers)},
  pages={1762--1777},
  year={2023}
}

@misc{li2025agenticragdeepreasoning,
      title={Towards Agentic RAG with Deep Reasoning: A Survey of RAG-Reasoning Systems in LLMs}, 
      author={Yangning Li and Weizhi Zhang and Yuyao Yang and Wei-Chieh Huang and Yaozu Wu and Junyu Luo and Yuanchen Bei and Henry Peng Zou and Xiao Luo and Yusheng Zhao and Chunkit Chan and Yankai Chen and Zhongfen Deng and Yinghui Li and Hai-Tao Zheng and Dongyuan Li and Renhe Jiang and Ming Zhang and Yangqiu Song and Philip S. Yu},
      year={2025},
      eprint={2507.09477},
      archivePrefix={arXiv},
      primaryClass={cs.CL},
      url={https://arxiv.org/abs/2507.09477}, 
}

@misc{li2026semanticsimilarityrethinkingretrieval,
      title={Beyond Semantic Similarity: Rethinking Retrieval for Agentic Search via Direct Corpus Interaction}, 
      author={Zhuofeng Li and Haoxiang Zhang and Cong Wei and Pan Lu and Ping Nie and Yi Lu and Yuyang Bai and Shangbin Feng and Hangxiao Zhu and Ming Zhong and Yuyu Zhang and Jianwen Xie and Yejin Choi and James Zou and Jiawei Han and Wenhu Chen and Jimmy Lin and Dongfu Jiang and Yu Zhang},
      year={2026},
      eprint={2605.05242},
      archivePrefix={arXiv},
      primaryClass={cs.IR},
      url={https://arxiv.org/abs/2605.05242}, 
}

\clearpage

\appendix
\section{Prompts and Tool Descriptions}
\label{app:prompts}

\subsection{LogicalRAG Search Tool}

\begin{promptbox}{LogicalRAG Search Tool Description (summary)}
Backend: BM25 sparse retrieval with Lucene query-string syntax.

Supported query controls:
  - Boolean logic: AND, OR, NOT, parentheses
  - Phrase matching: "exact phrase"
  - Field targeting: title:..., content:...
  - Term boosting: ^N
  - default_operator: OR or AND

Search strategy:
  - Start broad with distinctive keywords.
  - Narrow with AND, phrase matching, fields, or NOT when results are too broad.
  - Simplify or relax constraints when there are no hits.
  - For multi-hop questions, search one hop at a time.
  - Avoid repeated paraphrases that retrieve the same documents.

Examples:
  - broad keyword search
  - broad-to-narrow query refinement
  - distractor exclusion
  - exact title / phrase search
\end{promptbox}

\subsection{Agentic Hybrid Search Tool}

\begin{promptbox}{Agentic Hybrid Search Tool Description (summary)}
Backend: hybrid retrieval with BM25 keyword matching and FAISS dense retrieval,
merged by Reciprocal Rank Fusion (RRF).

Use case:
  - Robust retrieval when exact lexical overlap is incomplete.
  - Useful for paraphrases, aliases, indirect wording, and semantic matches.
  - Still benefits from explicit entity names, titles, years, locations, and rare relation terms.

Query format:
  - Use a short natural-language search query.
  - Do not use Boolean syntax such as AND, OR, NOT.
  - Prefer plausible semantic phrasing over loose keyword bags.
  - Unlike LogicalRAG, this tool does not expose logical query control;
    the agent controls retrieval through natural-language reformulation.

Search strategy:
  - Start with the most distinctive entity names and relation words.
  - For multi-hop questions, search one hop at a time.
  - If results are noisy, rewrite the query into a cleaner query for the current hop.
  - If direct wording fails, try aliases, paraphrases, adjectival forms,
    nominalizations, or related surface forms.
  - Increase max_results when more candidates around a relevant topic are needed.

Examples:
  - book title -> author
  - author -> education
  - direct wording -> paraphrased retry
  - focused topic -> increase max_results
\end{promptbox}

\FloatBarrier

\section{Experimental Settings}
\label{app:implementation}

Table~\ref{tab:implementation_details} summarizes the main retrieval and generation settings.
We use publicly available datasets, models, and software under their released terms and do not redistribute these artifacts.

\begin{table}[ht]
\centering
\scriptsize
\setlength{\tabcolsep}{5pt}
\renewcommand{\arraystretch}{1.05}
\begin{tabular}{p{0.30\linewidth}p{0.58\linewidth}}
\toprule
Component & Setting \\
\midrule
Sparse engine & OpenSearch \\
Indexed fields & \texttt{title}, \texttt{content} \\
Analyzer & OpenSearch default \\
BM25 & \(k_1=1.2,\ b=0.75\) \\
Hybrid fusion & RRF, \(K=60\) \\
Embedding model & All embedding-based components use Qwen3-Embedding-0.6B. \\
Query instruction & \texttt{Given a web search query, retrieve relevant passages that answer the query} \\
Medium FAISS & Flat index \\
KILT FAISS & IVFPQ \\
KILT FAISS params & \texttt{nprobe=64}, \texttt{hnsw\_m=32}, \texttt{pq\_m=64}, \texttt{pq\_nbits=8} \\
Max search turns & 8 \\
Agent model & Qwen3.5-Plus unless specified \\
Agent decoding & temp \(=0.6\), top-\(p=0.95\) \\
Judge model & Qwen3.5-Plus \\
Judge decoding & temp \(=0.3\) \\
Hardware & Intel Xeon Gold 6226R CPU, 4 RTX 3090 GPUs with 24GB memory each, 128GB RAM \\
\bottomrule
\end{tabular}
\vspace{1mm}
\caption{Experimental settings.}
\label{tab:implementation_details}
\end{table}

\section{Ablation Study}
\label{app:ablation}

We conducted two lightweight ablations on the first 400 examples of each KILT-scale dataset.

\paragraph{No logical operators.}
This variant disables \texttt{AND}, \texttt{OR}, and \texttt{NOT}, while still allowing keyword and phrase queries.

\paragraph{Syntax-only tool description.}
This variant keeps the query grammar but removes BM25-specific explanations and search strategies.

Table~\ref{tab:ablation_full} reports EM, F1, and Judge results; the overall trend is consistent across metrics.

\begin{table}[ht]
\centering
\scriptsize
\setlength{\tabcolsep}{8pt}
\renewcommand{\arraystretch}{1.04}
\begin{tabular}{llccc}
\toprule
Dataset & Variant & EM & F1 & Judge \\
\midrule
\multirow{3}{*}{2Wiki}
& No logical ops & 0.690 & 0.782 & 0.878 \\
& Syntax-only & 0.708 & 0.799 & 0.900 \\
& \textsc{LogicalRAG} & \textbf{0.710} & \textbf{0.803} & \textbf{0.907} \\
\midrule
\multirow{3}{*}{HotpotQA}
& No logical ops & 0.548 & 0.669 & 0.783 \\
& Syntax-only & 0.548 & \textbf{0.683} & 0.797 \\
& \textsc{LogicalRAG} & \textbf{0.555} & 0.679 & \textbf{0.800} \\
\midrule
\multirow{3}{*}{MuSiQue}
& No logical ops & \textbf{0.320} & 0.423 & 0.476 \\
& Syntax-only & 0.308 & \textbf{0.435} & 0.499 \\
& \textsc{LogicalRAG} & 0.315 & 0.427 & \textbf{0.501} \\
\midrule
Avg.
& No logical ops & 0.519 & 0.625 & 0.712 \\
& Syntax-only & 0.521 & 0.639 & 0.732 \\
& \textsc{LogicalRAG} & \textbf{0.527} & \textbf{0.636} & \textbf{0.736} \\
\bottomrule
\end{tabular}
\caption{
Full ablation results on the first 400 examples of each KILT-scale dataset.
}
\label{tab:ablation_full}
\end{table}

\section{Example Query Trajectory}
\label{app:trajectory_example}

\begin{tcolorbox}[
  enhanced,
  breakable,
  colback=promptbg,
  colframe=promptframe,
  coltitle=white,
  colbacktitle=prompttitle,
  title={Example LogicalRAG Trajectory},
  fonttitle=\bfseries,
  boxrule=0.6pt,
  arc=1mm,
  left=1.5mm,
  right=1.5mm,
  top=1mm,
  bottom=1mm
]
\scriptsize
\textbf{Question.}
What composer with an Italian libretto was born 4 March 1678?

\medskip
\textbf{Plan.}
Identify the person born on 4 March 1678, verify that the person is a composer, and confirm the Italian-libretto constraint.

\medskip
\textbf{Search 1.}
\texttt{born 4 March 1678}, \texttt{default\_operator=AND}.

\textbf{Key evidence.}
A timeline passage states that ``Italian Baroque composer Antonio Vivaldi is born on 4 March.''
The \textit{Antonio Vivaldi} page further states that ``Antonio Lucio Vivaldi ... 4 March 1678--28 July 1741'' was an Italian Baroque composer.

\medskip
\textbf{Search 2.}
\texttt{Antonio Vivaldi operas Italian libretto}, \texttt{default\_operator=OR}.

\textbf{Key evidence.}
Retrieved passages include \textit{Griselda (Vivaldi)}, which states that the opera uses an Italian libretto, along with other Vivaldi opera pages involving librettos.

\medskip
\textbf{Answer.}
\textit{Antonio Vivaldi}. The gold answer is \textit{Antonio Lucio Vivaldi}; LLM-as-a-Judge marks the answer as correct.
\end{tcolorbox}

\section{Model Scaling Details}
\label{app:model_scaling}

\definecolor{avggray}{RGB}{245,245,245}
\definecolor{gapgray}{RGB}{238,242,245}

\begin{table}[ht]
\centering
\scriptsize
\setlength{\tabcolsep}{3.5pt}
\renewcommand{\arraystretch}{1.04}
\begin{tabular}{llccccc}
\toprule
Model & Method & 2Wiki & Hotpot & MuSiQue & Avg. & Gap \\
\midrule
4B
& Hybrid  & 0.823 & 0.707 & 0.334 & 0.621 & -- \\
& Logical & 0.788 & 0.677 & 0.308 & 0.591 & -0.030 \\
\midrule
9B
& Hybrid  & 0.834 & 0.724 & 0.360 & 0.639 & -- \\
& Logical & 0.816 & 0.728 & 0.361 & 0.635 & -0.004 \\
\midrule
35B-A3B
& Hybrid  & 0.867 & 0.777 & 0.439 & 0.694 & -- \\
& Logical & 0.848 & 0.761 & 0.404 & 0.671 & -0.023 \\
\midrule
Plus
& Hybrid  & 0.882 & 0.794 & 0.471 & 0.716 & -- \\
& Logical & 0.886 & 0.784 & 0.481 & 0.717 & +0.001 \\
\bottomrule
\end{tabular}
\caption{
Model scaling results with LLM-as-a-Judge accuracy.
Hybrid denotes Agentic Hybrid, Logical denotes \textsc{LogicalRAG}, and Gap is Logical minus Hybrid on Avg.
}
\label{tab:model_scaling_judge}
\end{table}

\begin{table}[ht]
\centering
\scriptsize
\setlength{\tabcolsep}{3.0pt}
\renewcommand{\arraystretch}{1.03}
\begin{tabular}{lllccccc}
\toprule
Model & Method & Metric & 2Wiki & Hotpot & MuSiQue & Avg. & Gap \\
\midrule
4B
& Hybrid  & EM & 0.604 & 0.427 & 0.205 & 0.412 & -- \\
& Hybrid  & F1 & 0.707 & 0.559 & 0.306 & 0.524 & -- \\
& Logical & EM & 0.586 & 0.425 & 0.194 & 0.402 & -0.010 \\
& Logical & F1 & 0.682 & 0.544 & 0.286 & 0.504 & -0.020 \\
\midrule
9B
& Hybrid  & EM & 0.592 & 0.431 & 0.224 & 0.416 & -- \\
& Hybrid  & F1 & 0.693 & 0.574 & 0.334 & 0.534 & -- \\
& Logical & EM & 0.578 & 0.436 & 0.221 & 0.412 & -0.004 \\
& Logical & F1 & 0.682 & 0.570 & 0.326 & 0.526 & -0.008 \\
\midrule
35B-A3B
& Hybrid  & EM & 0.659 & 0.479 & 0.263 & 0.467 & -- \\
& Hybrid  & F1 & 0.754 & 0.619 & 0.379 & 0.584 & -- \\
& Logical & EM & 0.629 & 0.489 & 0.249 & 0.456 & -0.011 \\
& Logical & F1 & 0.728 & 0.622 & 0.359 & 0.570 & -0.014 \\
\midrule
Plus
& Hybrid  & EM & 0.713 & 0.534 & 0.302 & 0.516 & -- \\
& Hybrid  & F1 & 0.797 & 0.668 & 0.420 & 0.628 & -- \\
& Logical & EM & 0.701 & 0.524 & 0.302 & 0.509 & -0.007 \\
& Logical & F1 & 0.787 & 0.664 & 0.422 & 0.624 & -0.004 \\
\bottomrule
\end{tabular}
\caption{
EM and F1 model scaling results on the KILT-scale datasets.
Gap is Logical minus Hybrid on Avg.
}
\label{tab:model_scaling_emf1}
\end{table}

\section{Trajectory Metric Details}
\label{app:trajectory_metrics}

We computed the trajectory metrics in Figure~\ref{fig:trajectory_motivation} from retrieval actions in agent trajectories.
For each trajectory, we first grouped retrieval turns by their underlying search intent.
The grouping was produced by an LLM judge using only the issued retrieval queries, not the final answer or gold evidence.
Queries were assigned to the same intent if they attempted to retrieve the same missing fact or evidence, even when their surface forms differed.
We then applied deterministic post-processing to ensure that every retrieval turn was assigned to exactly one intent group.

We also judged whether each retrieval turn succeeded.
A turn was considered successful if at least one of its top-\(K\) retrieved passages was judged relevant to the issued query.
Only this binary success label was used when computing recovery.

\paragraph{Same-intent overlap.}
For each intent group with at least two retrieval turns, we computed the overlap between adjacent top-\(K\) retrieval results as the fraction of shared retrieved passage identifiers.
We then averaged these adjacent overlaps within the group.
The reported same-intent overlap is the macro average over all repeated-intent groups.
Lower overlap indicates that query revisions more effectively change the retrieved evidence for the same underlying information need.

\paragraph{Intent recovery.}
Intent recovery measures whether later query revisions can repair an initially failed search intent.
For each intent group, we checked whether the first retrieval attempt succeeded and whether any retrieval attempt in the same group eventually succeeded.
A group was considered recoverable only when it contained at least two retrieval turns and its first attempt failed.
It was considered recovered if any later attempt in the same group succeeded.

The intent recovery rate is the fraction of recoverable groups that were recovered.
Thus, singleton intents and intents that succeeded on the first attempt were excluded from the denominator.
This metric captures whether the agent can recover from an initial retrieval failure for the same search intent, rather than simply rewarding trajectories that issue more retrieval calls.

\end{document}